\documentstyle[aps]{revtex}

\begin{document}

\begin{center}
{\bf Radiation reaction 4-force: orthogonal or parallel to the 4-velocity?}

{\bf C\u alin Galeriu}

Department of Physics \\ Clark University \\ Worcester, MA, 01610, USA \\ cgaleriu@clarku.edu
\end{center}

{\bf Abstract}

In this note we point to some problems related to the standard derivation of the radiation
reaction 4-force, and we propose a new expression for this 4-force, parallel to the 
4-velocity.

\

The problem of whether a 4-force $F_{\mu}$ can be not orthogonal to the 4-velocity $v_{\mu}$ has appeared long ago, when the ponderomotive 4-force, in a system which dissipates energy by Joule heating, was considered \cite{pauli}. Abraham \cite{abraham} has shown that, since an inertial mass must be ascribed to every kind of energy, the rest mass $m_o$ of the system has to decrease, corresponding to the Joule heat dissipated. The ponderomotive 4-force must thus have a component parallel to the 4-velocity, and the equation of motion is modified accordingly:
\begin{equation}
F_{\mu} = {{\rm d} \over {\rm d} \tau} (m_o v_{\mu}) = m_o {{\rm d} v_{\mu} \over {\rm d} \tau} + v_{\mu} {{\rm d} m_o \over {\rm d} \tau}.
\end{equation}

The rate of energy dissipation, reflected in the variation of the rest mass, is given by
\begin{equation}
F_{\mu} v^{\mu} = - c^2 {{\rm d} m_o \over {\rm d} \tau} = - \gamma(v) {{\rm d} E \over {\rm d} t}.
\end{equation}

It is questionable why the same approach has not been applied to the case of radiation damping.
The radiative reaction force is introduced in order to satisfy an energy balance, for the nonrelativistic situation first \cite{jackson}. Thus the work done by the radiative reaction force has to equal the energy dissipated through electromagnetic radiation:
\begin{equation}
\int_{t_1}^{t_2} {\bf F} \cdot {\bf v} {\rm d} t = - {2 \over 3} {q^2 \over c^3} \int_{t_1}^{t_2} {\bf \dot v} \cdot {\bf \dot v} {\rm d} t \Rightarrow {2 \over 3} {q^2 \over c^3} \int_{t_1}^{t_2} {\bf \ddot v} \cdot {\bf v} {\rm d} t.
\end{equation}

The last part of (3) results from integrating by parts the Larmor power formula, if we assume that the motion is either periodic, or ${\bf \dot v} \cdot {\bf v} = 0$ at the moments $t_1$ and $t_2$. The radiation reaction force extracted this way is thus somehow averaged, and does not reflect the instantaneous damping force.

In a first questionable step, from (3) the radiation reaction force is extracted \cite{jackson} as
\begin{equation}
{\bf F} = {2 \over 3} {q^2 \over c^3} {\bf \ddot v}.
\end{equation}

We have to warn that, since in (3) ${\bf F}$ is in scalar product with ${\bf v}$, the only meaningful information that can be extracted is about the component of the force which is parallel to the velocity, $({\bf F} \cdot {\bf v}) {\bf v} / v^2$!
Another problem related to expression (4) is that it is not clear whether this force is indeed a {\it damping} force, pointing in the opposite direction than the velocity. This problem is evident if we consider the 'runaway' solution \cite{jackson}, in which the velocity, the acceleration and the acceleration's derivative are all parallel, pointing in the same direction, and increasing exponentially. This solution can be eliminated, but with the price of introducing acausal effects \cite{dirac}.

The force from (4) is generalized \cite{pauli} to the relativistic case by introducing the derivative with respect to the proper time $\tau$. 
\begin{equation}
F_{\mu} = {2 \over 3} {q^2 \over c^3} {{\rm d}^2 v_{\mu} \over {\rm d} \tau^2}.
\end{equation}

In a second questionable step an extra term, specifically needed to ensure the orthogonality between the 4-force and the 4-velocity, is added. The relativistic 4-force becomes:
\begin{equation}
F_{\mu} = {2 \over 3} {q^2 \over c^3} ({{\rm d}^2 v_{\mu} \over {\rm d} \tau^2} - {1 \over c^2} {{\rm d} v_{\nu} \over {\rm d} \tau} {{\rm d} v^{\nu} \over {\rm d} \tau} v_{\mu}).
\end{equation}

Since the only reason for being of the radiation reaction force is to account for the dissipation of energy, and this dissipation might be correlated to a decrease in the rest mass of the system, and furthermore only the component of the force parallel to the velocity enters the energy balance equation, we can safely consider the radiation reaction 4-force as being parallel to the 4-velocity. In other words, we make the intuitive assumption that a force parallel to the velocity will generalize to a 4-force parallel to the 4-velocity. We extract form (5) the component parallel to the 4-velocity, but pointing into the opposite direction, sought to describe the radiation reaction 4-force:
\begin{equation}
F_{\mu} 
= {2 \over 3} {q^2 \over c^3} {{\rm d}^2 v_{\nu} \over {\rm d} \tau^2} v^{\nu} {v_{\mu} \over - c^2}
= - {2 \over 3} {q^2 \over c^5} {{\rm d} v_{\nu} \over {\rm d} \tau} {{\rm d} v^{\nu} \over {\rm d} \tau} v_{\mu}.
\end{equation}

It is clear that this dissipative force (7) changes direction under time reversal, since the velocity changes direction. From (2) and (7) we can calculate the rate of energy dissipation:
\begin{equation}
{{\rm d} E \over {\rm d} t} = {- 1 \over \gamma(v)} F_{\mu} v^{\mu} = {- 1 \over \gamma(v)} {2 \over 3} {q^2 \over c^3} {{\rm d} v_{\nu} \over {\rm d} \tau} {{\rm d} v^{\nu} \over {\rm d} \tau}.
\end{equation}

In the nonrelativistic limit we recover the exact (not averaged!) Larmor power formula:
\begin{equation}
{{\rm d} E \over {\rm d} t} = - {2 \over 3} {q^2 \over c^3} {\bf \dot v} \cdot {\bf \dot v}.
\end{equation}

The force (7) also satisfies Dirac's \cite{dirac} relativistic energy-momentum balance equation:
\begin{equation}
{q^2 \over 2 \epsilon} {{\rm d} v_{\mu} \over {\rm d} \tau} - q v_{\nu} \textsf{F}_{\mu\ {\rm in}}^{\ \nu} - {2 \over 3} {q^2 \over c^3} ({{\rm d}^2 v_{\mu} \over {\rm d} \tau^2} - {1 \over c^2} {{\rm d} v_{\nu} \over {\rm d} \tau} {{\rm d} v^{\nu} \over {\rm d} \tau} v_{\mu}) = {{\rm d} B_{\mu} \over {\rm d} \tau}.
\end{equation} 

As Dirac pointed out, from this equation the radiation reaction force is not uniquely derived, but is determined up to a perfect differential $B_{\mu}$, subject only to the condition
\begin{equation}
{{\rm d} B_{\mu} \over {\rm d} \tau} v^{\mu} = 0.
\end{equation}

The solution (5) is the result of choosing 
\begin{equation}
B_{\mu} = ({q^2 \over 2 \epsilon} - m_o) v_{\mu},
\end{equation}
and of assuming that the rest mass $m_o$ is constant. The solution (6) is obtained for 
\begin{equation}
B_{\mu} = ({q^2 \over 2 \epsilon} - m_o) v_{\mu} - {2 q^2 \over 3 c^3} {{\rm d} v_{\mu} \over {\rm d} \tau},
\end{equation}
with a variable rest mass $m_o$ (but still with a constant charge $q$). The condition (11), using (2), reduces to (8).

While a few problems related to the 4-force (6) have been avoided with the solution (7), it is still unclear what is the mechanism by which the charged particle acquires rest mass (energy) from the field, such that it doesn't vanish through radiation. This problem, however, is also present in the classical solution, where it is supposed that "changes in the acceleration energy correspond to a reversible form of emission or absorption of field energy, which never gets very far from the electron" \cite{dirac}.

\end{document}